\documentclass{article}
\usepackage{cite}
\usepackage[utf8]{inputenc}
\usepackage{xcolor}
\usepackage{hyperref} 
\hypersetup{
    colorlinks=true,                
    breaklinks=true,                
    urlcolor= blue,
    linkcolor= black,                
    bookmarksopen=false,
    filecolor=black,
    citecolor=blue,
    linkbordercolor=blue
}
\AtBeginDocument{\hypersetup{pdfborder={0 0 0}}}

\usepackage{spconf,amsmath,graphicx}
\usepackage{xcolor}
\usepackage{tabularx}
\usepackage{booktabs}
\usepackage{tabularray}
\usepackage{xcolor}
\usepackage{multirow}
\usepackage{amsmath}
\usepackage{enumitem}
\usepackage{caption}
\usepackage{url}
\usepackage[T1]{fontenc}
\usepackage{subfig}
\usepackage{makecell}

\setlist[itemize]{itemsep=0.08em, parsep=0.08em}

\usepackage[flushleft]{threeparttable}

\newcommand{\blue}[1]{\textcolor{blue}{#1}}

\makeatletter
\renewcommand{\section}{\@startsection
  {section}%
  {1}%
  {}%
  {0.2\baselineskip}%
  {0.1\baselineskip}%
  {}
}%

\renewcommand{\subsection}{\@startsection
  {subsection}%
  {2}%
  {}%
  {0.15\baselineskip}%
  {0.11\baselineskip}%
  {}}%

\renewcommand{\subsubsection}{\@startsection
  {subsubsection}%
  {3}%
  {}%
  {0.1\baselineskip}%
  {0.1\baselineskip}%
  {}}%

\title{ENHANCING SPEAKER DIARIZATION WITH LARGE LANGUAGE MODELS:\\A CONTEXTUAL BEAM SEARCH APPROACH}
\name{Tae Jin Park, Kunal Dhawan, Nithin Koluguri, Jagadeesh Balam}
\address{
  NVIDIA, Santa Clara, USA
  }

\begin{document}
\ninept

\setlength{\abovedisplayskip}{6pt}
\setlength{\belowdisplayskip}{6pt}

\maketitle
\begin{abstract}
Large language models (LLMs) have shown great promise for capturing contextual information in natural language processing tasks. We propose a novel approach to speaker diarization that incorporates the prowess of LLMs to exploit contextual cues in human dialogues. Our method builds upon an acoustic-based speaker diarization system by adding lexical information from an LLM in the inference stage. We model the multi-modal decoding process probabilistically and perform joint acoustic and lexical beam search to incorporate cues from both modalities: audio and text. Our experiments demonstrate that infusing lexical knowledge from the LLM into an acoustics-only diarization system improves overall speaker-attributed word error rate (SA-WER). The experimental results show that LLMs can provide complementary information to acoustic models for the speaker diarization task via proposed beam search decoding approach showing up to 39.8\% relative  delta-SA-WER improvement from the baseline system. Thus, we substantiate that the proposed technique is able to exploit contextual information that is inaccessible to acoustics-only systems which is represented by speaker embeddings. In addition, these findings point to the potential of using LLMs to improve speaker diarization and other speech processing tasks by capturing semantic and contextual cues.
\end{abstract}
\section{Introduction}
\label{sec:intro}

Multi-speaker speech recognition has been often approached by applying speaker diarization system on the input audio and feeding the speaker homogeneous segments \cite{park2022review} to automatic speech recognition (ASR) systems \cite{medennikov2020target,cornell2023chime} or speaker diarization and speech recognition are done simultaneously \cite{shafey2019joint, kanda2022transcribe}. As we can observe from the most popular previous studies~\cite{park2022review}, lexical information is only infused to improve single speaker ASR based on beam search decoding \cite{ney1987data,scheidl2018word} which reduces word error rate (WER). In general, the probability of the next word or next token is calculated by n-gram or neural language model (LM) trained on ample amount of data and the probability of the next token or word is added to the probability of the token from the acoustic model to integrate the lexical cue from the trained language model and acoustic model. This type of beam search technique can be applied  to ASR models trained using connectionist temporal classification (CTC) loss \cite{graves2006connectionist} or recurrent neural network transducer (RNN-T) \cite{graves2012sequence}. 
\begin{figure}[t]
\centering
\includegraphics[width=0.48\textwidth]{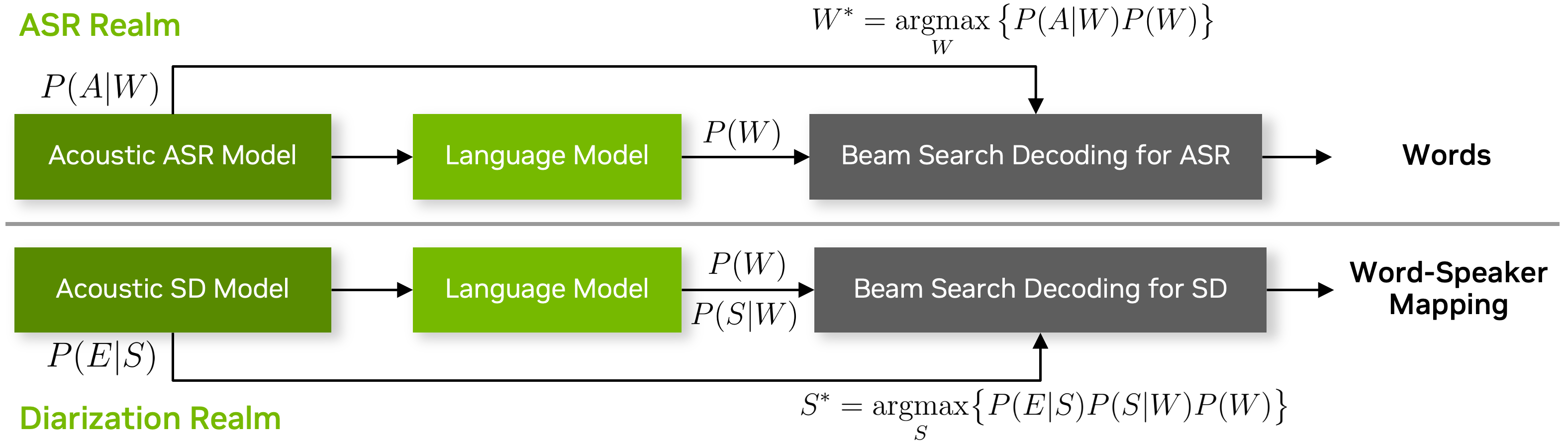}
\caption{The concepts of beam search decoding in the context of ASR and Speaker Diarization (SD).}
\label{fig:bsd_two_worlds}
\vspace{-4px}
\end{figure}

Despite the efficiency of the end-to-end ASR models which utilize an RNN-T architecture that integrates an internal LM, there are still significant benefits to be gained by these ASR models from incorporating an external language model. This improvement is mainly attributable to the disparities in the scale of training data available for acoustic and lexical modalities.~Specifically, datasets for training end-to-end ASR models are limited to those containing both audio and its corresponding transcript while language models can leverage text-only datasets, which are considerably larger and more diverse. In a similar vein, when it comes to speaker diarization, the volume of text data available is orders of magnitude greater than the volume of speaker-annotated audio data, especially when measured in terms of word count. Consequently, there is a potential for improvement by integrating the language models to enhance the performance of the speaker diarization task. 

The aim of this paper is to introduce an application of language models to the realm of speaker diarization, demonstrating the benefit of language model trained on large amount of text-only datasets. Fig.~\ref{fig:bsd_two_worlds} provides a comparative overview of LM applications in ASR realm and speaker diarization realm. We refer to our proposed technique as \textit{contextual beam search}, as it seeks the most probable word-speaker mapping $S$ by considering context from both modalities.

The utilization of lexical cues to speaker diarization, speaker turn detection and segmentation has been investigated for a long time yet remains less popular than acoustic-only speaker diarization research. Some of the earliest studies on this topic include the systems presented in \cite{canseco2004speaker,canseco2005comparative}, which leveraged linguistic patterns to identify speakers during the diarization process. Numerous studies have improved speaker segmentation or clustering accuracy by integrating ASR output to leverage lexical cues \cite{park2018multimodal, xia2022turn, khare2022asr}.\;Additionally, by merging speaker turn probabilities based on audio and text during the clustering phase \cite{park2020speaker}, lexical cues are further infused into speaker diarization results. Conversely, ASR and speaker diarization results have been jointly optimized to harness the lexical cues from ASR word outputs \cite{shafey2019joint, kanda2022transcribe}. More recently, the study presented in \cite{cheng2023exploring} introduced semantic information through neural embeddings generated by a spoken language processing (SLP) unit. Subsequently, a multimodal (audio-text) speaker change detector was proposed \cite{jung2023encoder}, along with the introduction of a speaker error correction (SEC) system \cite{paturi2023lexical} based on a pre-trained language model.

Our proposed method has the following distinction against the previous studies.~Firstly, our approach leverages a general-purpose LLM that is trainable on text-only datasets. This effectively addresses the data-sparsity challenge commonly associated with speaker diarization. In contrast, the systems proposed in the aforementioned studies \cite{shafey2019joint, xia2022turn, khare2022asr, park2020speaker, cheng2023exploring, jung2023encoder, paturi2023lexical} employ neural layers, such as RNN-T or transformer architectures to produce the final speaker logits and thus necessitate training on paired audio-text datasets, which our method circumvents. Secondly, our approach is not constrained by the number of speakers in a speaker diarization module as the decoding process does not rely on logits from a neural network layer. This is another advantage that systems in \cite{shafey2019joint, khare2022asr} cannot offer, as they are limited to a fixed number of speakers. Lastly, our approach functions similarly to general-purpose LMs work for end-to-end ASR models where an arbitrary LLM can be plugged in to improve the performance of the acoustic-only speaker diarization model. This offers significant advantages in various practical scenarios, especially when there is a need for modifying only the ASR model or the LM. For instance, when deploying the model for a different language, we can simply substitute ASR models and LLMs while using the same acoustic-only diarization model. 

\begin{figure}[t]
\centering
\includegraphics[width=0.44\textwidth]{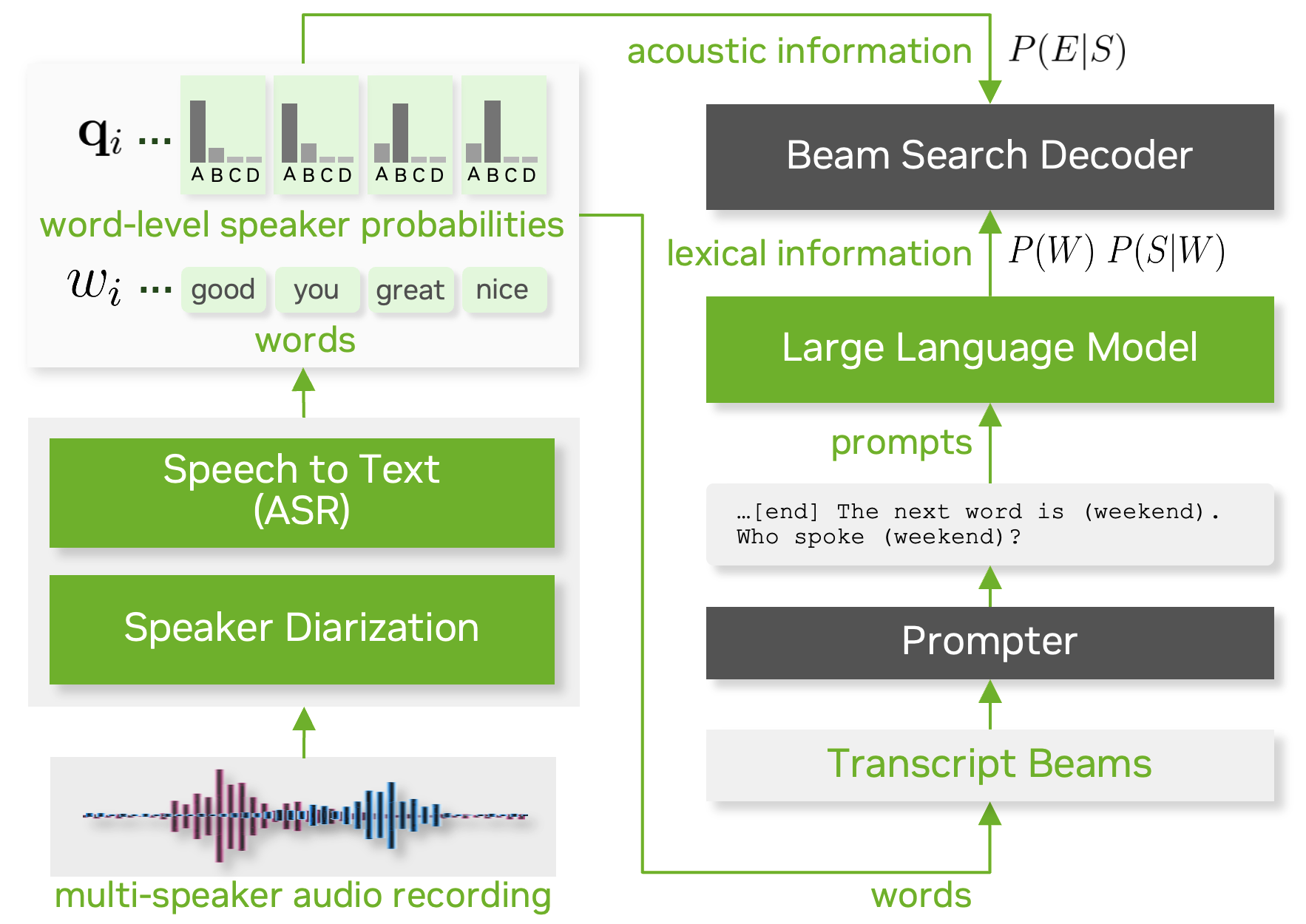}
\caption{Dataflow diagram of the proposed system.}
\vspace{-6px}
\label{fig:llm_ast_diar}
\end{figure}

\section{Probablistic Modeling for Beam Search}

\label{sec:format}
\subsection{Probablistic Formulations of ASR and Speaker Diarization}
In ASR frameworks, the task of converting speech-to-text (STT) revolves around building a model that translates a sequence of acoustic observations into a corresponding sequence of words. Formally, if we let $S$ denote the speaker identity, $W$ the word token, and $A$ the acoustic observation, the STT task can be mathematically represented as estimating the most likely word sequence $W$ given an acoustic observation $A$. This probability can be denoted as $P(W|A)$. Using Bayes' theorem, this can be represented as:
\begin{align}
W^{*} &= \operatorname*{argmax}_{W} \big\{P(W|A) \big\} \label{eq:lm_for_asr_1}\\ 
      &= \operatorname*{argmax}_{W} \Bigg\{\frac{P(A|W)P(W)}{P(A)} \Bigg\} \label{eq:lm_for_asr_2}\\ 
      &= \operatorname*{argmax}_{W} \big\{ P(A|W)P(W) \big\}, \label{eq:lm_for_asr_3}
\end{align}
where $W$ is a word or a token and $A$ is an acoustic observation. Expanding this idea to the realm of speaker diarization, our goal is to estimate the speaker label $S$ given both the acoustic observation $E$ and word $W$. Formally, we can express this as:
\begin{align}
S^{*} &=\operatorname*{argmax}_{S} \{ P(S|E,W) \} \label{eq:lm_for_sd_1} \\
      &=\operatorname*{argmax}_{S} \Bigg\{ \frac{P(E,W|S)P(S)}{P(E,W)} \Bigg\}  \label{eq:lm_for_sd_2}\\
      &=\operatorname*{argmax}_{S} \{ P(E,W|S)P(S) \} \label{eq:lm_for_sd_3}.
\end{align}
For the sake of simplifying our computations and model, we make an assumption of conditional independence between the acoustic observation $E$ and word $W$, given the speaker identity $S$. This assumption is mathematically represented as:
\begin{align}
P(E,W|S)&\overset{\mathrm{C.I.}}{=}P(E|S)P(W|S).
\label{eq:cond_indep}
\end{align}
From this conditional independence assumption, we can restructure Eq.~(\ref{eq:lm_for_sd_1}) to remove the unrelated term $P(W)$, leading to the essential expressions that require computation. The derivation is presented as:
\begin{align}
S^{*} &\overset{\mathrm{C.I.}}{=}\operatorname*{argmax}_{S} \bigl\{ P(E|S)P(W|S)P(S) \bigr\} \label{eq:sd_lm_with_ci_1}\\
      &=\operatorname*{argmax}_{S} \bigl\{ P(E|S)P(S|W)P(W) \bigr\}. \label{eq:sd_lm_with_ci_2}
\end{align}
In alignment with the ASR framework illustrated in Fig.~\ref{fig:bsd_two_worlds}, where $P(A|W)$ is represented as acoustic model, we utilize the acoustic-only diarization model to represent $P(E|S)$. Additionally, we derive $P(S|W)$ using pre-trained general-purpose language models, such as n-gram language models and LLMs.
It is crucial to differentiate between $P(W)$ in Eq.~(\ref{eq:lm_for_sd_3}) and $P(W)$ in Eq.~(\ref{eq:lm_for_asr_3}) because  the value of $P(W)$ in Eq.~(\ref{eq:lm_for_sd_3}) is contingent upon the condition that speakers are assigned to each word in the preceding word sequence. Let $w_i$ denote $i$-th word: each $i$ index corresponds to a specific word and its assigned speaker identity $k$, as illustrated in Fig.~\ref{fig:llm_ast_diar}.\;The given relationship can be expressed as:
\begin{align}
\mathbf{W}_{\mathbf{S}}^{C-1} = \{(w_{1}, \mathbf{q}_1),(w_{2}, \mathbf{q}_2), \cdots, (w_{C-1}, \mathbf{q}_{C-1})  \},
\label{eq:context_W_c}
\end{align}
where $C$ denotes the word sequence length (\textit{i.e.}, context length) and $q_i$ is the $i$-th speaker probability vector generated by the speaker diarization model. The matrix $\mathbf{S}$ encompasses a sequence of speaker probability vectors $\mathbf{q}$, represented as $\mathbf{S} = [\mathbf{q}_1, \mathbf{q}_2, \ldots, \mathbf{q}_{C-1}]$. With the notation $\mathbf{W}_{\mathbf{S}}^{C-1}$, the probability of a word given the past transcription $P(W)$ is expressed as:
\begin{align}
P(W) &= P(w_i|\textbf{W}_{\textbf{S}}^{C-1}).
\label{eq:p_w_def}
\end{align}
\subsection{Acoustic Inference: Speaker Diarization}
\label{sec:acoustic_infer}
We employ an improved version of Multi-scale Diarization Decoder (MSDD) model \cite{park2022multi}, which is introduced in \cite{park2023chime}.\;Given a maximum speaker limit $N_S$, the diarization output is manifested by $N_{S}$-dimensional floating point numbers, each representing the probability associated with each frame of 0.05 seconds in length. We utilize time-stamps derived from an ASR model to sample the diarization these diarization logit values. The corresponding speaker probability for $k$-th speaker can be described as 
\begin{align}
q_{k} = P(E|S)|_{S=k} = \frac{\sum_{t=1}^{T} p(S=k|E, t)}{\sum_{k}^{N_{S}}\sum_{t=1}^{T} p(S=k|E, t)},
\label{eq:word_diar_logits}
\end{align}
where $P(S|E, t)$ denotes the sigmoid logit value at time $t$ and $T$ denotes the number of diarization logit frames within the word. Consequently, we obtain an $N_{S}$-dimensional floating point probability vector $\mathbf{q}$ that sums up to one. As a result, we convert sigmoid values to probability values as we treat $P(E|S)$ as a probability measure.\;Note that any type of diarization or speaker recognition system can be employed to determine the speaker probability $P(E|S)$ for the sequence of $\mathbf{q}_i$ values, where $i$ denotes word index.\;After the diarization and ASR processes, or any other transcription and speaker recognition methods, we obtain a word sequence, $\mathbf{w}$ and corresponding speaker probability values per word denoted as $\mathbf{q}_i=[q_{1}, q_{2}, \dots,  q_{k}]_{i}$, as illustrated in Fig.~\ref{fig:llm_ast_diar}.

\subsection{Scoring for Beam Search Decoding}
\begin{figure}[t]
\centering
\includegraphics[width=0.46\textwidth]{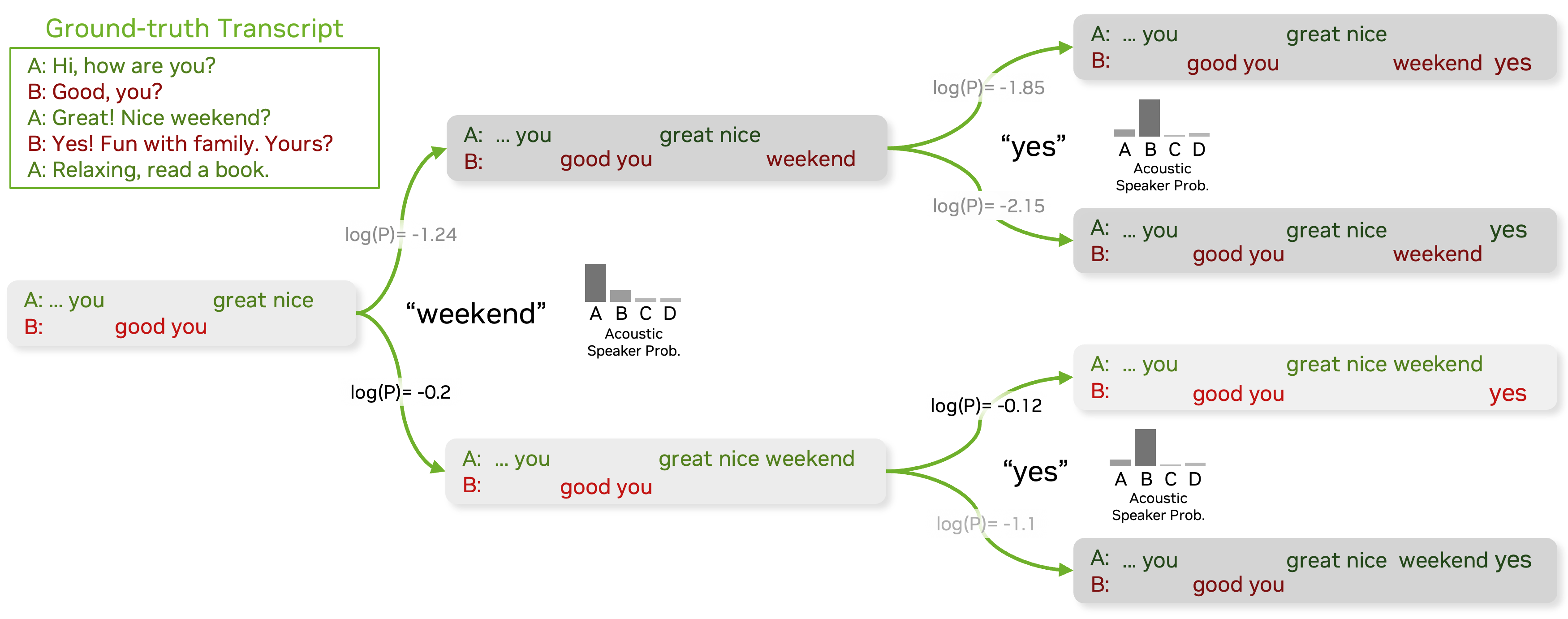}
\caption{Illustration of beam search for a two-speaker dialogue.}
\label{fig:bsd_example}
\vspace{-5px}
\end{figure}

Our proposed beam search decoding approach is based on Eq.~(\ref{eq:sd_lm_with_ci_2}). Let $\beta$ represent the mixing coefficient between acoustics-only speaker diarization and the language model, and let $\alpha$ represent the scaling parameter for $P(W)$. The beam search score function can then be formulated as:
\begin{align}
P_{BSD}(S) &= \log \bigl( P(E|S)\bigr) + \beta \log \bigl(P(S|W)P(W)^\alpha \bigr). 
\label{eq:bsd_log_prob}
\end{align}
We employ a modified version of the beam search decoder from \cite{pyctcdecode} which is capable of applying Eq.~(\ref{eq:bsd_log_prob}) as a score for beam search.  Fig.~\ref{fig:bsd_example} provides a visual representation of the beam search decoding process. The calculations of $P(E|S)$ and $P(S|W)$ will be covered in section~\ref{sec:lexical_infer}. The role of the term $P(W)$ is crucial. Since the sum of $P(S|W)$ for all speakers $S$ equals 1, if all speakers have an equal probability in a lexical sense, $P(S|W)$ becomes $1/N_{S}$. For instance, if we consider cases where multiple speakers utter the same filler words, we are likely to see $P(S|W)$ approaches $1/N_{S}$ as there is not significant lexical context to distinguish one's speaker identity from another. The term $P(W)$ addresses these situations by assigning a relatively low probability, compensating for the uncertain lexical cue. Hence, the beam search predominantly relies on $P(E|S)$ to select the most probable speaker for the corresponding word. 

While $P(W)$ can serve as a confidence parameter for lexical context, the degree of compensation can be controlled by the parameter $\alpha$. From a mathematical perspective, if $\alpha$ is close to 0, we give less importance on the confidence of the language model and as $\alpha$ increases, we further suppress the lexical context, proportional to the word probability.

\section{Lexical Inference}
\label{sec:lexical_infer}
We employ two types of language model for comparison: the KenLM\cite{heafield2011kenlm} based n-gram language model and  the GPT-based LLM from the Megatron-LM framework in the NeMo toolkit \cite{shoeybi2019megatron, kuchaiev2019nemo}. 
\subsection{Baseline n-gram language model}
\subsubsection{Speaker probability from n-gram}
We use a 4-gram language model \cite{panayotov2015librispeech} that is publicly available and has approximately 8M probability value entries. The following example demonstrates speaker-wise transcript and the \textit{next word} that should be appended to a speaker's last word.

\vspace{-4px}
\noindent\rule{\linewidth}{0.8pt}
{\scriptsize 
\begin{itemize}[label={},leftmargin=*,labelsep=0.0em,topsep=0pt]
\vspace{-1px}
\item \textbf{[Speaker-0]}\;{\ttfamily \textless s\textgreater\;how are you doing these days\textless /s\textgreater \textless s\textgreater\;well tell me more\textless /s\textgreater }
\item \textbf{[Speaker-1]}\;{\ttfamily \textless s\textgreater\;things are going very well\textless /s\textgreater\textless s\textgreater\;there is a project that i'm }
\item \textbf{[All-Speakers]}\;{\ttfamily \textless s\textgreater\;how are you doing these days\textless /s\textgreater\textless s\textgreater\;things are going very well\textless /s\textgreater  \textless s\textgreater well tell me more\textless /s\textgreater \textless s\textgreater\;there is a project that i'm }
  \item \textbf{[Next Word]} {\ttfamily working}
\vspace{-4px}
\end{itemize} 
}
\vspace{-2px}
\noindent\rule{\linewidth}{0.8pt}
For the n-gram language model, we use the start-of-sentence (SOS) token, \textless s\textgreater~and the end-of-sentence (EOS) token, \textless /s\textgreater. These tokens can significantly influence the probability of the n-gram score for a given word sequence. The following equation describes the probability of having $w_{\text{next}}$ given that all $C$ words assigned with the particular speakers in $S$:
\begin{align}
P_{S}\left( w_{i}=w_{\text{next}} \right) 
& = P(w_{S, i}|\mathbf{W}_{S}^{C-1}) \label{eq:n_gram_spk_a1}\\ 
& =P(S, W), \label{eq:n_gram_spk_a2}
\end{align}
where $\textbf{W}_{S}^{C-1}$ is a word sequence $w_{1}, \ldots, w_{n}$ for the given speaker $S$, and $C-1$ represents the length of the context window for speaker $k$. Therefore,  the probability for the $k$-th speaker among $N_{S}$ speakers is denoted by the following equation:
\begin{align}
P(S|W)\rvert_{S=k} &= \frac{P(S, W)}{P(W)}\bigg|_{S=k} = \frac{P(w_{k, i}|\mathbf{W}_{k}^{C-1})}{\sum_{k=1}^{N} P(w_{k, i}|\mathbf{W}_{k}^{C-1})} \label{eq:n_gram_spk_b1}.
\end{align}
\subsubsection{Word Probability}
As we discussed in the previous section, we calculate the $P(W)$ term separately. Note that this probability differs from $P(W)$ in Eq.~(\ref{eq:n_gram_spk_b1}) since we use the entire word sequence, which includes all speakers, to calculate the $P(W)$ value using the following equation:
\begin{align}
P(W) & =\sum_{k=1}^{N_S} P(w_{k, i}|\mathbf{W}_{k}^{L-1})\rvert_{w_{k,i}=w_{\text{k, next}}} \\
& =P(w_{\text{next}}|\mathbf{W}_{S}^{L-1}),
\end{align}
where $L$ is the length of the context that includes all $N$-speakers.

\subsection{Speaker Diarization Prompt for LLM}
While the baseline n-gram language model works for the purpose of estimating the most probable speaker, a neural language model trained on large amount of data can parse more context in the given text-based conversation. This allows it to estimate a more accurate speaker probability based on lexical cues.

\begin{table*}[ht]
\centering
\setlength{\tabcolsep}{4pt}
\begin{tabular}[t]{r|cc|cc|cc|cc|cc||c|cccc}
\Xhline{3\arrayrulewidth}
Language & \multicolumn{10}{c||}{\footnotesize (\;LM for $P(S|W)$, LM for $P(W)\;)$} &\multirow{3}{*}{\text{ASR}} & \multicolumn{4}{c}{\multirow{3}{*}{\text{Speaker Diarization}}} \\\cline{2-11}
Model (LM) & \multicolumn{2}{c|}{\text{-}} & \multicolumn{2}{c|}{\text{{\footnotesize All n-gram}}} & \multicolumn{2}{c|}{\text{{\footnotesize(LLM,\;n-gram)}}} & \multicolumn{2}{c|}{\text{{\footnotesize(n-gram,\;LLM)}}} & \multicolumn{2}{c||}{\text{{\footnotesize All LLM}}} &  & \multicolumn{4}{c}{\text{}} \\\cline{0-10}
\text{Matching} & \multicolumn{2}{c|}{\text{\footnotesize TS-match}} & \multicolumn{2}{c|}{\text{\footnotesize BSD}} & \multicolumn{2}{c|}{\text{\footnotesize BSD}} & \multicolumn{2}{c|}{\text{\footnotesize BSD}} & \multicolumn{2}{c||}{\text{\footnotesize BSD}} & &&& &  \\
\hline
\text{Metric} & $\Delta$SA & $\Delta$cp & $\Delta$SA & $\Delta$cp & $\Delta$SA & $\Delta$cp & $\Delta$SA & $\Delta$cp & $\Delta$SA & $\Delta$cp  & \text{WER} & Miss & FA & CER & DER  \\
\Xhline{3\arrayrulewidth}
\text{CH-others}   & 8.45 & 8.28 & 8.07 & 7.86 & 8.36 & 8.18 & 8.61 & 8.41 & \textbf{7.76} & \textbf{7.58} & 25.15 & 5.97  & 9.79 & 6.04 & 21.80  \\
\text{CH-109}      & 5.05 & 4.97 & 3.70 & 3.64 & 3.70 & 3.63 & 4.42 & 4.30 & \textbf{3.24} & \textbf{3.16} & 23.07 & 5.55  & 4.39 & 1.44 & 11.38 \\
\text{AMI-MH-dev}  & 8.22 & 8.20 & 3.47 & 3.45 & 3.49 & 3.47 & 3.52 & 3.50 & \textbf{3.45} & \textbf{3.42} & 23.68 & 15.89 & 4.02 & 1.02 & 15.89  \\
\text{AMI-MH-test} & 6.38 & 6.41 & 3.92 & 3.88 & \textbf{3.74} & \textbf{3.72} & 3.98 & 3.94 & 3.84 & 3.80 & 23.25 & 12.96 & 4.98 & 1.36 & 19.03  \\
\Xhline{3\arrayrulewidth}
\end{tabular}
\vspace{-2px}
\caption{\footnotesize Performance evaluation of multi-speaker ASR based on time-stamp (TS) matching and beam search decoding (BSD). }\
\vspace{-31px}
\label{table:er_table}
\end{table*}

\subsubsection{Speaker probability from LLM}
In the LLM-based calculation of $P(S|W)$, we rely on a prompt that asks the expected speaker the expected speaker to provide the subsequent word. An illustrative example of such a prompt for the subsequent word {\ttfamily working} is provided below.
\vspace{-5px}
\noindent\rule{\linewidth}{0.8pt}
\begin{itemize}[label={},leftmargin=*,labelsep=0.0em,itemsep=0.01pt,partopsep=0.01pt]
\vspace{-1px}
{\scriptsize {\ttfamily
  \item \lbrack Speaker0\rbrack:\,how are you doing these days 
  \item \lbrack Speaker1\rbrack:\,things are going very well 
  \item \lbrack Speaker0\rbrack:\,well tell me more 
  \item \lbrack Speaker1\rbrack:\,there is a project that i'm 
  \item \lbrack end\rbrack \vspace{-1.0px}
  \item Question:~The next word is~(working).\;Who spoke (working)?  \vspace{-11.0px}
  \item Answer:[Speaker\blue{1]}}}
\vspace{-10px}
\end{itemize}
\noindent\rule{\linewidth}{0.8pt}
Using the provided prompt template, we simulate the probability $P(S, W)$ by sampling the probability values from the token indicating the speaker index that follows the token labeled as {\ttfamily speaker}. The prompt includes the text leading up to {\ttfamily speaker}. From this, we can derive $P(S|W)$ using the subsequent equation:
\begin{align}
P(S|W)\rvert_{S=k} &= \frac{p_k}{\sum_{k=1}^{N_S} p_{k}}  \label{eq:llm_spk_a1},
\end{align}
where $p_k$ denotes the simulated probability calculated from the logit values in the LLM output for speaker index $k$. It is worth noting that the n-gram approach inherently cannot account for interactions between two speakers when determining speaker probability.\;In contrast, the LLM approach, the LLM approach allows for the consideration of interplay and combined information between speakers when estimating the speaker probability $P(S|W)$.
\subsubsection{Word Probability}
In case of calculating $P(W)$, we use a specific input to sample the probability of the last word. We employ the same dialogue prompt as used for the speaker probability. However, we insert the term \textit{next word} and remove any text following the {\ttfamily \lbrack end\rbrack} token. With this prompt, the word probability can be expressed as:
\begin{align}
P(W)\rvert_{W=w_{\text{next}}} &= \frac{p_{d}}{\sum_{d=1}^{D} p_{d}}, \label{eq:llm_word_b1}
\end{align}
where $d$ and $D$ represent the token index and the total number of tokens, respectively.
\section{Experimental results}
\subsection{Experiment Settings}
\subsubsection{Pre-trained Models}
\begin{itemize}[leftmargin=*,labelsep=0.05em]
    \item \textbf{ASR:} We employ a Conformer-CTC model~\cite{gulati2020conformer} that is implemented using the NeMo Toolkit~\cite{kuchaiev2019nemo}. The ASR model has approximately 122M parameters and 1024 tokens. 
    \item \textbf{Speaker Diarization:} We employ an improved version~\cite{park2023chime} of the MSDD model~\cite{park2022multi}, which has 32M parameters. It builds upon Titanet~\cite{koluguri2022titanet}, Transformer Encoder~\cite{vaswani2017attention}, and Clustering~\cite{park2019auto}.
    \item \textbf{LLM\footnote{Details regarding model training and datasets will be provided in \cite{nvidia2023nemo}}:} We use an LLM based on~\cite{shoeybi2019megatron} which is a scaled version of GPT~\cite{radford2019language} model of 2B parameter size. Trained on a 1.1T token subset of the dataset curated by NVIDIA Data Collector (NDC) \cite{nvidia2023nemo}, containing 70\% of English dataset, 15\% of Stack v1.1 dataset \cite{huggingfaceBigcodethestackdedupDatasets} and 15\% of multi-lingual datasets from Common Crawl~\cite{luccioni2021s}.
\end{itemize}
\subsubsection{Evaluation Metric}

\begin{itemize}[leftmargin=*,labelsep=0.05em]
\item \textbf{WER}: The WER is determined using a hypothesis script generated from a single-channel mixed audio clip and a transcript that contains words in onset order. Note that this WER should be differentiated from the channel-specific or speaker-specific WER in other studies. We use text normalization tool from \cite{openai_whisper} for evaluation.
\item \textbf{SA-WER}: The Speaker-Attributed WER (SA-WER)~\cite{cornell2023chime} is a WER metric grounded in speaker mapping, as established by speaker diarization. SA-WER measures the WER by comparing hypothesis and reference transcripts for each specific speaker mapping. 
\item \textbf{cpWER}: The concatenated minimum-permutation
word error rate (cpWER) \cite{watanabe2020chime} is a metric designed to capture both ASR and diarization accuracy.
cpWER is calculated by taking the minimum WER from concatenated transcripts of multiple speakers across all potential permutations.
\item \textbf{$\Delta$cp} and \textbf{$\Delta\text{SA}$}: delta-cpWER and delta-SA-WER follows the following relationship.
\begin{align*}
\vspace{-4px}
\Delta\text{cp} &= \text{cpWER} - \text{WER} \\
\Delta\text{SA} &= \text{SA-WER} - \text{WER}
\end{align*}
Our evaluation is based on the assumption that $\Delta$cp and $\Delta\text{SA}$ reflect the diarization error in cpWER and SA-WER, respectively.
\end{itemize}
\subsection{Datasets and Systems Overview}
Table~\ref{table:er_table} shows the performance for each setup and dataset. For AMI-MH (Mixed Headset)~\cite{kraaij2005ami} dataset, we use only-word version \cite{landini2022bayesian} of annotation. The Call Home American English Speech (CHAES, LDC97S42) is a corpus composed solely of English telephonic speech data. CH-109 is a subset that includes two speakers per session, whereas CH-others signifies the other sessions found within CHAES. We optimize $\alpha$, $\beta$ in Eq.~(\ref{eq:bsd_log_prob}) and context window length $C$ in Eq.~(\ref{eq:context_W_c}), beam-width for beam search decoding on CH-others and AMI-MH-dev then use the parameters for CH-109 and AMI-MH-test, respectively. For this parameter optimization, we employ Optuna~\cite{akiba2019optuna}. The Diarization Error Rate (DER) is computed using a collar of 0.25 seconds while including overlaps.

In terms of systems, \textit{TS-match} is a system that relies on time-stamps (TS) to match speaker diarization time-stamps with the decoded word time-stamps from ASR system. \textit{All n-gram} and \textit{All LLM} are systems where speaker probability and word probability values are calculated from the n-gram LM and the LLM, respectively. Conversely, \textit{(LLM,\;n-gram)} and \textit{(n-gram,\;LLM)} employ one model for speaker probability and another for word probability, in accordance with the provided notation. All experiments involving the LLM were conducted using the NVIDIA TESLA V100 GPU.

\subsection{Analysis of Experimental Results}
In Table~\ref{table:er_table}, the performance of TS-matching is compared with four combinartory setups that integrate both n-gram and LLM. Regarding $\Delta$cp and $\Delta$SA, it is crucial to highlight that speaker confusion resulting from speaker diarization results in a double error count because it manifests as an insertion for one speaker and a deletion for another. The proposed method improves the baseline system's delta-SA-WER by up to 39.8\%.
We can deduce couple of findings: Firstly, the trend shows that applying same type of language model leads to a lower error rate. This is likely because the $P(W)$ and $P(S|W)$ are less prone to discrepancies. Secondly, LLM seems more effective in estimating $P(S|W)$, given that the average performance of applying LLM for $P(S|W)$ demonstrates superior performance over n-gram.  We speculate that the performance difference stems from the fact that the n-gram model considers only a single speaker, while LLM processes the entire transcription, taking into account all speakers within the context window. This contextual understanding is a distinct advantage LLM holds over n-gram LM, providing a more nuanced estimation of the speaker. However, using LLM demands roughly 15 times more computational time during inference when compared to the n-gram LM. This discrepancy underscores the potential need for a trade-off, suggesting a hybrid \textit{(LLM, n-gram)} configuration.

\section{Conclusions}
In this paper, we introduce a beam search decoding-based approach for applying a language model to speaker diarization. The proposed method offers a key advantage: it uses individually optimized models for ASR, diarization, and LLM. By training each model independently, we can leverage large-scale data sources specific to each domain. Moreover, even without fine-tuning the ASR, diarization, and LLM models, the proposed method achieves significant improvement over conventional approaches. This methodology can be seamlessly adapted to accommodate multilingual contexts by integrating multilingual ASR and LLM models—areas that have seen significant advancements recently. Looking forward, our future research will focus on several areas: Firstly, we intend to integrate beam search decoding for ASR and diarization by applying a single LLM which can achieve a more efficient system with enhanced accuracy. Secondly, we plan to integrate the ASR and diarization decoders to obtain more accurate timestamps alongside speaker logits, aiming for a streamlined multi-speaker ASR. Finally, we will explore ways to improve the model by introducing more sophisticated context, either by fine-tuning or prompt-tuning the LLM using domain-specific data.
\clearpage

\vfill
\bibliographystyle{IEEEbib}
\bibliography{strings,refs}

\end{document}